\documentstyle[aps,prd,tighten,epsf]{revtex}
\begin{document}
\draft
\newcommand{\be}{\begin{equation}}
\newcommand{\ee}{\end{equation}}
\newcommand{\ben}{\begin{eqnarray}
\displaystyle}
\newcommand{\een}{\end{eqnarray}}
\newcommand{\la}{{\lambda}}
\newcommand{\ta}{{\tilde a}}
\newcommand{\si}{{\sigma}}
\newcommand{\th}{{\theta}}
\newcommand{\C}{{\cal C}}
\newcommand{\D}{{\cal D}}
\newcommand{\cA}{{\cal A}}
\newcommand{\cT}{{\cal T}}
\newcommand{\cO}{{\cal O}}
\newcommand{\eeo}{\cO ({1 \over E})}
\newcommand{\G}{{\cal G}}
\newcommand{\cL}{{\cal L}}
\newcommand{\T}{{\cal T}}
\newcommand{\M}{{\cal M}}
\newcommand{\p}{\partial}
\newcommand{\na}{\nabla}
\newcommand{\ssum}{\sum\limits_{i = 1}^3}
\newcommand{\dssum}{\sum\limits_{i = 1}^2}
\newcommand{\tal}{{\tilde \alpha}}
\newcommand{\tp}{{\tilde \phi}}
\newcommand{\tim}{{\tilde \mu}}
\newcommand{\tr}{{\tilde \rho}}
\newcommand{\tir}{{\tilde r}}
\newcommand{\rp}{r_{+}}
\newcommand{\hr}{{\hat r}}
\newcommand{\rv}{{r_{v}}}
\newcommand{\dr}{{d \over d \hr}}
\newcommand{\dR}{{d \over d R}}
\newcommand{\hhf}{{\hat \phi}}
\newcommand{\hhM}{{\hat M}}
\newcommand{\hhQ}{{\hat Q}}
\newcommand{\hht}{{\hat t}}
\newcommand{\hhr}{{\hat r}}
\newcommand{\hhS}{{\hat \Sigma}}
\newcommand{\hhD}{{\hat \Delta}}
\newcommand{\hhm}{{\hat \mu}}
\newcommand{\hro}{{\hat \rho}}
\newcommand{\hhz}{{\hat z}}
\newcommand{\tD}{{\tilde D}}
\newcommand{\tB}{{\tilde B}}
\newcommand{\hT}{\hat T}
\newcommand{\tF}{\tilde F}
\newcommand{\tT}{\tilde T}
\newcommand{\hC}{\hat C}
\newcommand{\ep}{\epsilon}
\newcommand{\bep}{\bar \epsilon}
\newcommand{\ppp}{\varphi}
\newcommand{\Ga}{\Gamma}
\newcommand{\ga}{\gamma}
\newcommand{\hth}{\hat \theta}
\title{The question of Abelian Higgs hair expulsion from extremal
dilaton black holes}

\author{ Rafa{\l} Moderski}
\address{JILA, University of Colorado \protect \\
CB 440, Boulder, CO 80309-0440 \protect \\
and \protect \\
Nicolaus Copernicus Astronomical Center \protect \\
Polish Academy of Sciences \protect \\
00-716 Warsaw, Bartycka 18, Poland \protect \\
moderski@camk.edu.pl }

\author{Marek Rogatko}
\address{Technical University of Lublin \protect \\
20-618 Lublin, Nadbystrzycka 40, Poland \protect \\
rogat@tytan.umcs.lublin.pl \protect \\
rogat@akropolis.pol.lublin.pl}
\date{\today}
\maketitle
\smallskip
\bigskip
\begin{abstract}
It has been argued that the extremal dilaton black holes exhibit a flux
expulsion of Abelian-Higgs vortices. We re-examine carefully the
problem and give analytic proofs for the flux expulsion always takes
place. We also conduct numerical analysis of the problem using three
initial data sets on the horizon of an extreme dilatonic black hole,
namely, core, vacuum and sinusoidal initial conditions. We
also show
that an Abelian-Higgs vortex can end on the extremal dilaton black
hole. Concluding, we calculate the backreaction of the Abelian-Higgs
vortex on the geometry of the extremal black hole and
drew a conclusion that a straight cosmic string and the extreme dilatonic
black hole hardly knew their presence.
\end{abstract}
\baselineskip=18pt
\par
\section{Introduction}
Wheeler's metaphoric dictum {\it black holes have no hair}, motivated by
the earlier investigations on uniqueness theorems for black holes
(see Ref.\cite{bl} and references therein), has
had a great influence on the development of black hole physics. In the
recent years there has been a considerable resurgance of mathematical
works on black hole equilibrium sates. We are faced with the discoveries
of black hole solutions in many theories in which Einstein equations are
coupled with self-interacting matter fields. The discovery of Bartnik
and McKinnon \cite{ba} of a nontrivial point like structure in
Einstein-Yang-Mills system reveals new realms of nontrivial solutions
to Einstein-non-Abelian gauge systems.
\par
Recently, the other sort of problems has been taken into account.
The {\it no hair } conjecture has been
extended to the problems when
topologies of some field configurations are not trivial. 
The story began with the work of
Aryal {\it et
al.} \cite{vi}.
Though they considered a pure gravity theory, they wrote down
the metric of a black hole with a conical deficit and claimed
that this was 
the metric describing the Schwarzschild black hole
with a cosmic string passing through it. 
This work gave rise to the researches concerning the extension of the
afore mentioned {\it no hair} conjecture.
In 
Ref.\cite{tl}, the extension of these
considerations announced
the existence of the Euclidean Einstein equations
corresponding to a vortex sitting on the horizon of the black hole.
Achucarro {\it et al.} \cite{gg} presented the numerical and
analytic evidence for an Abelian-Higgs vortex can act as a long hair
for the Schwarzschild black hole. Chamblin {\it et al.} \cite{cha}
generalized
the analysis of Ref.\cite{gg} and allowed the black hole to be
charged. Using numerical analysis they found that all of the fields
connected with the vortex were expelled from the extremal black hole.
The extremal black hole displayed an analog of a {\it Meissner effect}.
However, the recent works of Bonjuor {\it et al.}\cite{bon}
show subtleties in
the treatment of the event horizon, showing that a flux expulsion can 
occur but it does not do so in all cases. Analytic proofs for an
expulsion and a penetration of a flux in the case of the extremal charged
black hole were also presented.
\par
Nowadays, it seems that the superstring theories are the most promissing
candidates for a consistent quantum theory of gravity.
Numerical studies of the solutions to the
low-energy string theory, i.e., the Einstein-dilaton black holes in
the presence of a Gauss-Bonnet type term, disclosed that they were
endowed with a nontrivial dilaton hair \cite{ka}. This dilaton hair is
express
ed in term of its ADM mass \cite{adm}. The extended moduli and
dilaton hair and their associated axions for a Kerr-Newmann black hole
background were computed in Ref.\cite{ka1}. 
On the other hand, a full analysis of a cosmic string in dilaton
gravity was given in Ref.\cite{san}.
An Abelian
Higgs vortex in the background of an Euclidean electrically charged
dilaton
black hole was studied in Ref.\cite{mr}. It was shown that this kind
of the Euclidean black hole can support a vortex solution on a horizon
of the black hole. The vortex effect was to cut out a slice out of the
considered black hole geometry.
In Ref.\cite{mr1} the authors argued that an 
electrically charged dilaton black
hole could support a long range field of a Nielsen-Olesen string. Using
both numerical and perturbative techniques the properties of an
Abelian-Higgs vortex in the presence of the considered black hole were
investigated. In the case of an extreme dilatonic black hole the analog
of the {\it Meissner effect} was revealed.
\par
In this paper we will try to provide some continuity with our previous
works \cite{mr}-\cite{mr1}. Namely, we shall
re-examine the problem of the flux
expulsion in the presence of the extremal dilaton black hole in the
light of the arguments recently quoted by Bonjour {\it et al.}\cite{bon}.
We have provided analytical proofs that vortices will wrap around the
extreme dilatonic black hole. Then, we conduct the numerical analysis of
the problem.
\par
The paper is organized as follows. In Sec.II we briefly review some
results obtained in Ref.\cite{mr1}
that will be needed, so that the paper becomes
self-contained. We also conduct analytic considerations of the
expulsion problem of the Nielsen-Olesen vortex in the presence of an
extreme dilatonic black hole. In Sec.III, we give a numerical analysis of
the problems taking into account three initial data sets on the horizon
of the black hole, i.e., core, vacuum and sinusoidal initial
conditions. We also pay attention to the problem of a vortex which
terminates on the extremal dilaton black hole. Before concluding our
considerations, in Sec.IV, we discuss the gravitational backreaction.

\section{Nielsen-Olesen vortex and a dilaton black hole}
In this section we shall review some material published in
Ref.\cite{mr1} in
order to establish notation and convention and for the paper to be
self-contained. We also gave some analytic arguments in favour of a flux
expulsion.
In our consideration we shall study an Abelian-Higgs vortex in the
presence of a dilaton black hole assuming a complete separation of
degrees of freedom of each of the objects. Our system will be described
by the action which is the sum of the action for a low-energy
string theory \cite{gro}, which in the Einstein frame, has the form
\be
S_{1} = \int \sqrt{-g}d^4 x
\left [ {R\over 16 \pi G} - 2 ( \na \phi )^2 - e^{-2 \phi} F^2 \right ],
\ee
and $S_2$ is
the action for an Abelian Higgs system minimally coupled to gravity
and be subject to spontaneous symmetry
breaking. It yields
\be
S_{2} = \int \sqrt{-g}d^4 x
\left [ - D_{\mu} \Phi^{\dagger} D^{\mu} \Phi - {1 \over 4}
\tB_{\mu \nu} \tB^{\mu \nu} - {\la \over 4}
\big ( \Phi^{\dagger} \Phi - \eta^2 \big )^{2} \right ],
\ee
where $\Phi$ is a complex scalar field, $D_{\mu} = \na_{\mu} + ie
B_{\mu}$ is the gauge covariant derivative. $\tB_{\mu \nu}$ is the
field strength associated with $B_{\mu}$, while $F_{\alpha \beta} =
2 \na_{[ \alpha} A_{\beta ]}.$ 
As in Ref.\cite{gg} one can define the real
fields $X, P_{\mu}, \chi$ by the relations
\ben
\Phi (x^{\alpha}) = \eta X (x^{\alpha}) e^{i \chi (x^{\alpha})},\\ 
B_{\mu}(x^{\alpha}) = {1 \over e} \left [ P_{\mu}(x^{\alpha}) -
\na_{\mu} \chi (x^{\alpha}) \right ].
\een
Equations of motion derived from the action $S_{2}$ are given by
\ben
\na_{\mu} \na^{\mu} X - P_{\mu}P^{\mu} X - {\la \eta^2 \over 2}
\big ( X^2 - 1 ) X = 0, \\
\na_{\mu} \tB^{\mu \nu} - 2 e^2 \eta^2 X^2 P^{\nu} = 0. 
\een
The field $\chi$ is 
not a dynamical quantity. 
In a flat spacetime  the Nielsen-Olesen vortices have the cylindrically
symmetric solutions of the forms
\ben
\Phi = X (\rho) e^{i N \phi}, \\
P_{\phi} = N P(\rho),
\een
where $\rho$ is the cylinder radial coordinate, $N$ is the winding
number. 
\par
As far as
a static, spherically symmetric solution of the equations
of motion derived from the action $S_{1}$ is concerned
it is determined by the metric of a
charged dilaton black hole. The metric may be written as
\cite{db}
\be
ds^2 = - \left ( 1 - {2 M \over r} \right ) dt^2 +
{ d r^2 \over \left ( 1 - {2 M \over r} \right ) } + r \left ( 
r - {Q^2
\over M} \right ) (d \th
eta^2 + \sin^2 \theta  d \ppp^2),
\ee
where we define $r_{+} = 2M$ and $r_{-} = {Q^2 \over M}$ which
are related to the mass $M$ and charge $Q$ by the relation $Q^2 =
{r_{+} r_{-} \over 2} e^{2 \phi_{0}}$. The charge of the dilaton
black hole $Q$, couples to the field $F_{\alpha \beta}$, is unrelated
to the Abelian gauge field $B_{\mu \nu}$ associated with the
vortex. The dilaton field is given by $e^{2\phi} = \left ( 1 - {r_{-} \over r} \right )
e^{-2\phi_{0}}$, where $\phi_{0}$ is the
dilaton's value at $r \rightarrow \infty$. The event horizon is
located at $r = r_{+}$. For $r = r_{-}$ is another
singularity, one can however ignore it because $r_{-} < r_{+}$. The extremal black hole
occurs when $r_{-} = r_{+}$,
when $Q^2 = 2M^2 e^{2\phi_{0}}$.
\par
In what follows, we shall consistently assume that $X$ and $P_{\phi}$
are functions of $r, \theta$ coordinates. Then,
equations of motion
for these fields are given by
\ben \label{m}
{1 \over r (r - {Q^2 \over M})} \p_{r} \left [
\big ( r - {Q^2 \over M} \big ) (r - 2 M ) \p_{r} X \right ]
&+& {1 \over  r (r - {Q^2 \over M}) \sin \theta }
\p_{\theta} [ \sin \theta \p_{\theta} X ] -
{N^2 P^2
X \over  r (r - {Q^2 \over M}) \sin^2 \theta} -
{1 \over 2} X (X^2 -1 ) = 0,\\
\p_{r} \left [ \left ( 1 - {2 M \over r} \right ) \p_{r} P \right ] &+&
{\sin \theta \over  r (r - {Q^2 \over M})} \p_{\theta}
[csc \theta \p_{\theta} P ] - {X^2 P \over \beta
} = 0,
\label{mm}
\een
where $\beta = {\la \over 2 e^2} = m_{Higgs}^2 / m_{vect}^2$ is the
Bogomolnyi parameter.  When $ \beta \rightarrow \infty$, the Higgs
field decouples and like in the Reissner-Nordstr\"om case \cite{cha},
one can study a free Maxwell field in the electrically charged black
hole spacetime. The other situation will arise when $P = 1$. It will
be the case of a global string in the presence of the electrically
charged dilaton black hole.
\par
One can show Ref.\cite{gh}, that in normal spherically symmetric
coordinates $X$
is a function of
$\sqrt{g_{33}}$ and one
will try with the coordinates $R = \sqrt{r (r - {Q^2 \over M})} \sin
\theta$, namely $X = X(R)$ and $P_{\phi} = P_{\phi}(R)$. 
Taking into account {\it the thin string} limit,
i.e., $M \gg 1$, equations for $X(R)$ and $P(R)$ can be reduced to the
Nielsen-Olesen type up to the errors of an adequate 
order (see Ref.\cite{mr1}).
\par
The main task of our work will be to answer the question about the flux
expulsion in the case of the extremal dilaton black hole. Some numerical
arguments concerning the so-called {\it Meissner effect} were quoted in
Ref.\cite{cha}. Now, having in mind the arguments of Ref.
\cite{bon}, we examine this
problem carefully. We begin with the analytical considerations. From
now on, we shall consider only the extremal dilaton black hole, for which
$r_{+} = r_{-}$.
\par
As one can see from Eqs.(\ref{m}-\ref{mm}),
the flux expulsion solution $X = 0$, $ P = 1$ always solves these 
equations of
motion. Thus, our strategy will be to show the nonexistence of a
penetration solution. First of all, we assume that a piercing solution
does exist. This requirement is fulfilled when one has the nontrivial
solutions $X(\theta)$ and $ P(\theta)$ which is symmetric around 
$\theta =
{\pi \over 2}$. For this value of the angel $X$ has a maximum and $P$ a
minimum. Expanding equations of motion for $P(\theta)$ near the poles
indicates that $P_{,\theta} = 0$ at the poles.
Thus, there exists such a point $\theta_{0} \in (0, {\pi \over 2})$ for
which the second derivative of $P(\theta)$ is equal to zero, namely
$P_{,\theta  \theta}(\theta_{0}) = 0$ 
and $P_{,\theta}  (\theta_{0}) < 0$.
In the extremal black hole case Eq.(\ref{mm}) has the form as follows
\be
P_{, \theta \theta} - P_{, \theta} \cot \theta = 0.
\ee
Having in mind that $P_{, \theta \theta}(\theta_{0}) = 0$ and $ \cot
\theta_{0} \neq 0$ in the considered interval of $ \theta$, we reach to
the contradiction with our starting assumption that $P_{,
\theta}(\theta_{0}) < 0$. 
This is sufficient to conclude that
the flux expulsion must always take place.
\par
Further,
after proving that the flux expulsion must take place for a sufficiently
thick string, we proceed to the case of a thin string. As was remarked
in Ref.\cite{bon} 
in order to consider the case of a thin string one has to
look at the Eqs.(\ref{m}-\ref{mm}) in the exterior region of the
horizon. 
To begin with,
we assume that there is a flux expulsion. Therefore near the
horizon of the extremal dilaton black hole, one has
$\p_{r}[ (r - 2 M)^2 \p_{r}
X] > 0 $, $X^{3}2M (2M - {Q^2  \over M}) \ll 1$. Hence from Eq.
(\ref{m}) we obtain
\be
\sin^2 \theta \p_{r}[ (r - 2 M)^2 \p_{r} X ] + \sin \theta 
\p_{\theta} [ \sin \theta \p_{\theta} X ] - X N^2 P^2 = 0.
\label{ts}
\ee
The function $X$ is symmetric around ${\pi \over 2}$, having maximum
$X_{m}$. Integrating Eq.(\ref{ts}) on the interval $(\theta, {\pi \over
2})$, for $\theta > \beta$ we arrive at the inequality
\be
\p_{\theta} X(\theta) > X(\theta) \left [
{N^2 \over \sin \theta} \ln \tan  {\theta \over 2} \right ].
\ee
Then using
the fact that $X_{, \theta \theta} < 0$ on $[ \theta_{0},
{\pi \over 2}]$, one can deduce that $X_{, \theta} < {X (\theta) -
X (\theta_{0}) \over \theta - \theta_{0}} < {X(\theta) \over \theta -
\beta} $.
This shows that the inequality
\be
{1 \over N^2} > (\theta - \beta){1 \over \sin \theta} \ln \tan {\theta
\over 2},
\label{in}
\ee
must hold over the range $\theta \in (\beta, {\pi \over 2})$ for
the expulsion to occur.
From the graph of the function $\zeta (\theta) =
{1 \over \sin \theta} \ln \tan {\theta \over 2}$, we deduce that on the
interval $\theta \in (\beta, {\pi \over 2})$ 
one has
$\zeta (\theta) < 0$.
Then, the inequality (\ref{in}) always holds and we have the expulsion
of the vortex for the extremal dilaton black hole.
\par
The analytic solution of equations of motion for the dilaton black hole
which size was small comparing to the vortex size were considered by the
present authors in Ref.\cite{mr1}. It was done for the large $N$-limit.
Then it happened that,
inside a core of the vortex the gauge symmetry will be unbroken, and
the expectation that ${X^2 \over \beta} \approx 0$ is well justified.
These all provided the solution of Eq.(\ref{mm}) as
\be
P \approx 1 - p R \sin^2 \theta,
\label{p}
\ee
where $p$ is an integration constant equal to twice the magnetic field
strength at the center of the core \cite{gg}. 
On the other hand, the solution for $X$ yields
\be
X \approx b^{N} (r) \sin^{N} \theta.
\ee
where $b(r)$ is given by Eq.(21) in Ref.\cite{mr1}. The exact form of $X$
ensures its vanishing on the extremal black hole horizon.
Concluding we see that, using analytical arguments one has always the
expelling of the fields from the extreme dilaton black hole horizon.
As we see in the next section these analytical considerations are fully
confirmed by numerical investigations.

\section{Numerical results}
To confirm our results from previous section we performed number of
numerical calculations for extremal
dilaton black holes with strings. The numerical
method is simply the same as in our previous article \cite{mr1} (see
also Ref.\cite{pftv}).
Namely, the fields X
and P are replaced with their discreted values on the polar grid $(r, \theta)
\rightarrow (r_i= 2 M + i \Delta r, \enskip \theta_k=k \Delta \theta)$ 
according to the
difference version of the equations of motion (\ref{m}) and (\ref{mm})
\ben
\label{diff1}
X_{00}={\left ( 1-{2 M \over r} \right ) {X_{+0}+X_{-0} \over (\Delta r)^2}
+ {1 \over r (r - {Q^2 \over M})} {X_{0+}+X_{0-} \over (\Delta \theta)^2} +
\left [ {1 \over r} + {r -2M \over r (r - {Q^2 \over M})} \right ]
{X_{+0}-X_{-0} \over 2 \Delta r} + {\cot \theta \over r (r - {Q^2 \over
M})}{X_{0+}-X_{0-} \over 2 \Delta \theta} \over \left ( 1 - {2M \over r}
\right ) {2 \over (\Delta r)^2} + {2 \over r (r - {Q^2 \over M}) (\Delta
\theta)^2} + {1 \over r (r - {Q^2 \over M})}{\left ( {P_{00} N \over \sin
\theta} \right )}^2 + {1 \over 2} (X_{00}^2 - 1)} \\
P_{00}={\left ( 1-{2 M \over r} \right ) {P_{+0}+P_{-0} \over (\Delta r)^2}
+ {1 \over r (r - {Q^2 \over M})} {P_{0+}+P_{0-} \over (\Delta \theta)^2} +
{2M \over r} {P_{+0}-P_{-0} \over 2 \Delta r} - {\cot \theta \over r (r -
{Q^2 \over M})}{P_{0+}-P_{0-} \over 2 \Delta \theta} \over \left ( 1 - {2M
\over r} \right ) {2 \over (\Delta r)^2} + {2 \over r (r - {Q^2 \over M})
(\Delta \theta)^2} + {1 \over \beta} X_{00}^2},
\label{diff2}
\een
On the horizon we used 
\ben
\label{bound1}
X_{00}={{1 \over \Delta r} X_{+0} + {\cot \theta \over (2M - {Q^2 \over
M})}{X_{0+}-X_{0-} \over 2 \Delta \theta} + {1 \over (2M - {Q^2 \over
M})}{X_{0+} +X_{0-} \over (\Delta \theta)^2} \over {1 \over \Delta r} +
M(X_{00}^2 - 1) + {1 \over (2M - {Q^2 \over M})}{\left ( {P_{00} N \over
\sin \theta} \right )}^2 + {2 \over (2M - {Q^2 \over M}) (\Delta
\theta)^2}} \\
P_{00}={{1 \over \Delta r} P_{+0} - {\cot \theta \over (2M - {Q^2 \over
M})}{P_{0+}-P_{0-} \over 2 \Delta \theta} + {1 \over (2M - {Q^2 \over
M})}{P_{0+}+P_{0-} \over (\Delta \theta)^2} \over {1 \over \Delta r} + {1
\over \beta} 2 M X_{00}^2 + {2 \over (2M - {Q^2 \over M}) (\Delta
\theta)^2}}
\label{bound2}
\een
The boundary conditions are imposed on large radii and for a string core
according to
\be
(X, P) = \cases{ (1, 0), & $r \rightarrow \infty$
\cr (0, 1),& $r \ge 2M, \enskip \theta = 0, \pi.$}
\label{bound}
\ee
Boundary conditions on the horizon are guessed at the begining of
calculation and then updated in accordance with Eqs. (\ref{bound1}) and
(\ref{bound2}). The process of relaxation and 
updating of the fields on the horizon of the black hole
is repeated until convergence take place.
\par
As an initial guess we used three sets of different boundary conditions as
proposed by Bonjour~{\it et al.}~\cite{bon}, i.e.,
\begin{enumerate}
\item{core: $X = 0$, $P = 1$,}
\item{vacuum: $X = 1$, $P = 0$,}
\item{sinusoidal: $X(\theta) = \sin \theta$, $P = 1$.}
\end{enumerate}
In all above cases we received the same final configuration of the fields
always showing the string expulsion from the extreme
dilatonic black hole. These
confim our previous theoretical predictions. The results 
of the numerical calculations
are presented in
Fig.~1.
\par
Further, we pay our attention to the case of a cosmic string ending on
the extreme dilatonic black hole. This is an important configuration
as the main phenomenological input to the instantons mediating
defect decays (se, e.g., \cite{c1}-\cite{c2}). 
Namely, in Ref.\cite{gh} it was shown 
that the Nielsen-Olesen solution
could be used to construct regular metrics which represented vortices
ended on black holes either in a static equilibrium or
accelerating off to infinity. The latter metric depicts a cosmic string
which is eaten by accelerating black holes.

To consider numerically 
a string ending on the black hole we have to slightly
modify the boundary conditions (\ref{bound}). They remain the same at
the outer boundary and in the string core for $\theta=0$. For
$\theta=\pi$ we initially set $X=1$ and $P=0$. During the calculation
values of the fields were updated on each step according to $P_{,
\theta}=0$ and $X_{, \theta}=0$. These were chosen to assure the
numerical stability. On the horizon we set vaccum boundary conditions
to avoid the possibility that flux expulsion, if it occures, may be
caused by the inapropriate choice of initial conditions on the
horizon. The results of our simulations for extreme dilatonic
black hole with $M=1.0$ and
$\beta=N=1$ are presented in Fig.~2.

\section{Gravitational Backreaction}
The gravitational backreaction of the string on the dilaton black hole
geometry was studied in Ref.\cite{mr1}. As one can compare the resultant
metric we shall obtain and the conical metric gained in Ref.\cite{mr1}
they
are the same. But in our previuos attitude we did not pay attention to
the corrections of other fields in the theory. Therefore our
previous results were not correct. In this section we consider the
backreaction problem taking into account all the fields in the theory.
\par
In order to consider the gravitational backreaction effect of the
string superimposed on the dilaton black hole one needs to consider
a general static axially symmetric solution to the
Einstein-Maxwell-dilaton Abelian-Higgs equations of motion. First, we
find a coordinate transformation which enalbes us to write the {\it
spherical} dilaton black hole metric in the axisymmetric form.
Using the coordinate transformation as follows:
\ben
\rho^2 &=& (r - 2M) \left ( r - {Q^2  \over M} \right )
\sin^2 \theta, \\
z &=& \left ( r - M - {Q^2 \over 2M} \right ) \cos \theta,
\een
the metric of the charged dilaton black
hole may be rewritten in the axisymmetric form, namely
\be
ds^2 = - e^{2 \psi_{0}} dt^2 + \alpha_{0}^2 e^{- 2 \psi_{0}} d\ppp^2 + e^{-
2\psi_{0}  + 2
\gamma_{0}} (d \rho^2  + dz^2),
\label{zero}
\ee
where
\ben
e^{- 2 \psi_{0}} &=& {r \over r - 2M},\\
e^{- 2 \psi_{0} + 2 \gamma_{0}} &=& 
{\left ( r - {Q^2 \over M} \right ) r \over
A} ,\\
\rho^2 &=& {\alpha}_{0}^2, \\
A &=& (r - 2M) \left ( r - {Q^2  \over M} \right )
\cos^2 \theta + {\sin^2 \theta
 \over 4} \left [
(r - 2M) +  \left ( r - {Q^2  \over M} \right ) \right ]^2.
\een
The dilaton black hole metric (\ref{zero}) will constitute,
to the zeroth order, our background solution.
The relevant  equations of motion become
\be
\p_{\rho} \left ( e^{-2 \phi - 2 \psi} \alpha F_{0 \rho} \right )
+ \p_{z} \left ( e^{-2 \phi - 2 \psi} \alpha F_{0 z} \right ) = 0,
\label{f}
\ee
\be
\p_{\rho} \left ( \alpha \p_{\rho} \phi  \right ) +
\p_{z} \left ( \alpha \p_{z} \phi  \right ) + e^{-2 \phi}
\sqrt{- g} \left [ 
(F_{0 \rho})^2 + (F_{0 z})^2 \right ] = 0,
\ee
\be
\alpha_{, z z} + \alpha_{,\rho \rho} = \sqrt{- g}
\left ( \cT_{z}{}{}^{z} + \cT_{\rho}{}{}^{\rho} \right ),
\label{m1}
\ee
\be
( \alpha \psi_{, z} )_{, z} + ( \alpha \psi_{,\rho} )_{,\rho}
=  - {1 \over 2}  \sqrt{- g}
\left ( \cT_{0}{}{}^{0} -  \cT_{z}{}{}^{z} - 
\cT_{\rho}{}{}^{\rho} -  \cT_{\ppp}{}{}^{\ppp} 
\right ),
\label{m2}
\ee
\be
- \ga_{,\rho} (\alpha^2_{, \rho} + \alpha^2_{, z}) +
\alpha \alpha_{, \rho} (\psi^2_{, \rho} - \psi^{2}_{, z})  + 
2 
\alpha \alpha_{, z} \psi_{, \rho} \psi_{, z} + 
\alpha_{, \rho} \alpha_{,\rho \rho} + \alpha_{, z} \alpha_{,\rho
z} =  \sqrt{- g}
\left ( \alpha_{,\rho} \cT_{z}{}{}^{z} - 
\alpha_{, z} \cT_{z}{}{}^{\rho} \right ),
\label{m3}
\ee
\be
\ga_{, z} (\alpha^2_{, \rho} + \alpha^2_{, z}) -
\alpha \alpha_{, z} (\psi^2_{, \rho} - \psi^{2}_{, z})  -
2 \alpha \alpha_{, \rho} \psi_{, \rho} \psi_{, z}
+ \alpha_{, z} \alpha_{,\rho \rho} - \alpha_{,\rho} \alpha_{,\rho 
z} =  \sqrt{- g}
\left ( \alpha_{, z} \cT_{z}{}{}^{z} - 
\alpha_{, \rho} \cT_{z}{}{}^{\rho} \right ),
\label{m4}
\ee
\be
\ga_{, z z} + \ga_{, \rho \rho} + \psi^2_{, z} +
\psi^2_{,\rho} =  \sqrt{- g}
\left ( e^{2 \ga - 2 \psi} \cT_{\ppp}{}{}^{\ppp} \right ),
\label{m5}
\ee
where the energy momentum tensor $\cT_{\alpha}{}{}^{\beta}$ is given by
\be
\cT_{\alpha}{}{}^{\beta} = \ep T_{\alpha}{}{}^{\beta}(string) +
T_{\alpha}{}{}^{\beta}(F, \phi),
\ee
where $\ep = 8\pi G\eta^2$ which is assumed to be small. This
assumption is well justified because, e.g., for the
GUT string
$\ep \leq 10^{-6}$.
The first term is the contribution from the string and it has the
following explicit form:
\ben
T_{0}{}{}^{0}(string) &=& - e^{-2(\ga - \psi)} \left (
X^2_{,\rho} + X^2 _{, z} \right ) - {X^2 P^2 e^{2\psi} \over
\alpha^2} - {\beta \over \alpha^2} e^{-2\ga + 4\psi}
\left ( P^2_{, \rho} + P^2_{, z} \right ) - V(X), \\
T_{\ppp}{}{}^{\ppp}(string) 
&=& - e^{-2(\ga - \psi)} \left (
X^2_{,\rho} + X^2 _{, z} \right ) + {X^2 P^2 e^{2\psi} \over
\alpha^2} + {\beta \over \alpha^2} e^{-2\ga + 4\psi}
\left ( P^2_{, \rho} + P^2_{, z} \right ) - V(X), \\
T_{\rho}{}{}^{\rho}(string) &=& e^{-2(\ga - \psi)} \left (
X^2_{,\rho} - X^2 _{, z} \right ) - {X^2 P^2 e^{2\psi} \over
\alpha^2} + {\beta \over \alpha^2} e^{-2\ga + 4\psi}
\left ( P^2_{, \rho} - P^2_{, z} \right ) - V(X), \\
T_{z}{}{}^{z}(string) &=&  e^{-2(\ga - \psi)} \left (
X^2_{, z} - X^2 _{,\rho} \right ) - {X^2 P^2 e^{2\psi} \over
\alpha^2} + {\beta \over \alpha^2} e^{-2\ga + 4\psi}
\left ( P^2_{, z} - P^2_{,\rho} \right ) - V(X), \\
T_{\rho}{}{}^{z}(string) &=& 2 e^{-2(\ga - \psi)} \left (
X_{,\rho}X_{, z} + {\beta \over \alpha^2}
P_{, \rho}P_{, z} \right ), 
\een
where $V(X) = {(X^2 -1)^2 \over 4}$.
The electromagnetic dilaton contribution is given by
\be
T_{\mu \nu}(F, \phi) = e^{- 2 \phi} \left (
4 F_{\mu \rho} F_{\nu}{}{}^{\rho} - g_{\mu \nu}F^2 \right )
- 2 g_{\mu \nu}(\na \phi)^2 + \na_{\mu}\phi \na_{\nu} \phi.
\label{ff}
\ee
Having in mind Eq.(\ref{ff}) one notices that the
electromagnetic-dilaton energy momentum
tensor always fulfills
\be
T_{\rho}{}{}^{\rho}(F, \phi) + T_{z}{}{}^{z}(F, \phi) = 0.
\ee
As in Ref.\cite{gg} we will solve the Einstein-Maxwell-dilaton
equations 
iteratively, i.e., $\alpha = \alpha_{0} + \ep \alpha_{1} $ etc.
Following Ref.\cite{mr1} we can use the coordinates
$ R = \sqrt{r (r - {Q^2 \over M})} \sin \theta = \rho e^{-\psi_{0}},$ 
which yields that near the core of the string where 
$\sin \theta \approx \cO(M^{-1})$,
one gets $
R_{,\rho}^{2} + R_{, z}^{2} \sim e^{2\ga_{0} - 2\psi_{0}}.$
This relation implies that near the core
of the string, to the zeroth order, the energy momentum tensor  reads
as follows:
\ben
T_{(0) 0}^{}{}{0}(str
ing) &=& - V(X_{0}) - \left ( \dR X_{0} \right )^{2}
- {X_{0}^2 P_{0}^2 \over R^2} - {\beta \over R^2} \left ( \dR P_{0} \right )^2 , \\
T_{(0) \ppp}^{}{}{\ppp}(string) &=& 
- V(X_{0}) - \left ( \dR X_{0} \right )^{2}
+ {X_{0}^2 P_{0}^2 \over R^2} + {\beta \over R^2} \left ( \dR P_{0} \right )^2 , \\
T_{(0) \rho}^{}{}{\rho}(string) &=& 
- V(X_{0}) - \left ( \dR P_{0} \right )^{2} 
+ e^{-2(\ga_{0} - \psi_{0})} \left [ {\beta
\over R^2} \left ( \dR P_{0} \right )^2 + \left ( \dR X_{0} \right )^2 \right ]
(R_{,\rho}^2 - R_{,z}^2), \\ 
T_{(0) z}^{}{}{z}(string) &=& 
- V(X_{0}) - \left ( \dR P_{0} \right )^{2} - e^{-2(\ga_{0} - \psi_{0})} \left [
{\beta
\over R^2} \left ( \dR P_{0} \right )^2 + \left (\dR X_{0} \right )^2 \right ] 
(R_{,\rho}^2 - R_{,z}^2), \\
T_{(0) \rho}^{}{}{z}(string) &=&
2 e^{-2(\ga_{0} - \psi_{0
})} \left [{\beta \over R^2} 
\left ( \dR P_{0} \right )^2 + 
\left (\dR X_{0} \right )^2 \right ] R_{,\rho}R_{, z}.
\een
which is the purely function of $R$. As in 
the Schwarzschild case \cite{gg}, this
strongly suggests to look for the metric perturbat
ions as a function
of $R$.
\par
As in Ref.\cite{bon}
we assume that the first order perturbed solutions take form
\be
\alpha_{1} = \rho a (R), \qquad \psi_{1} = \psi_{1}(R), \qquad
\ga_{1} = \ga_{1}(R), \qquad \phi_{1} = \phi_{1}(R), \qquad
A_{\mu}{}{}^{(1)} = f(R) A_{\mu}{}{}^{(0)}.
\ee
Further, carry out the
computing of the necessary derivatives one gets the following 
equation for $a(R)$ 
\be
{d^2 \over dR^2}a(R) + {2\over R} \dR a(R) = T_{\rho}{}{}^{\rho}(string)
+ T_{z}{}{}^{z}(string) + \cO( M^{-2}).
\ee
From which one can reach to the following expression for $a(R)$:
\be
a(R) = \int_{R}^{\infty} {1 \over R^2} dR \int_{0}^{R} R'^2
\left ( - 2V(X_{0}) - {2X_{0}^2 P_{0}^2 \over R'^2} \right ) dR'.
\label{a}
\ee
Eq. (\ref{a}) can be rewritten as
\be
a(R) \sim - {A} + {B \over R},
\ee
where
\ben
A &=& \int_{0}^{R} R ( T_{(0) \rho}{}{}^{\rho}(string) +
T_{ (0) z}{}{}^{z}(string) ) dR, \\
B &=&  \int_{0}^{R} R ( T_{(0) \rho}{}{}^{\rho}(string) +
T_{(0)z}{}{}^{z} (string)) dR.
\een
On the other hand, Eq.(\ref{f}) implies
\be
{d^2 \over dR^2}f + {1 \over R^2} \dR f = {\rho^2 \over R^2 r^2}
\left [
\dR a - 2 \left ( \dR \phi_{1} + \dR \psi_{1} \right ) \right ]
= \cO(M^{-2}),
\ee
which yields that $f = f_{0}$ is equal to a constant value.
The magnetic correction one can get either directly or using the
duality transformation \cite{hor}, which implies
$F \rightarrow \ast F$, $\phi \rightarrow - \phi$, where $\ast F_{\mu \nu}
= {1 \over 2}e^{- 2 \phi} \ep_{\mu \nu \rho \delta} F^{\rho \delta}$.
Turning our attention to Eq.(\ref{m2}) and taking into account the
value of $f_{0}$, one finds for
$\psi_{1}$
\be
{d^2 \over dR^2}\psi_{1} + {1 \over R} \dR \psi_1 = 
- {1 \over 2}
\left ( T_{(0) 0}{}{}^{0} - T_{ (0) z}{}{}^{z} -  
T_{(0) \rho}{}{}^{\rho} -  T_{(0) \ppp}{}{}^{\ppp}
 \right ) (string) + {4 \rho^2 \over R^2} \left (
{Q^2 \over r^4} \right ) [f_{0} - (\psi_{1} + \phi_{1})].
\label{p1}
\ee
The result of the integration of
Eq.(\ref{p1}) is given by
\be
\psi_{1}(R) = - {1 \over 2}
\int_{R}^{\infty} {1 \over R} dR \int_{0}^{R} R'
\left (  2V(X_{0}) - {2 \beta \over R'^2} \left ( {dP_{0} \over dR}
\right )^2 \right ) dR'.
\ee
Consider now Eq.(\ref{m5}). For $\ga_{1}(R)$, one gets 
the expression of the form
\be
\ga_{1}(R) = \int_{R}^{\infty} dR \int_{0}^{R}
T_{\ppp}{}{}^{\ppp}(string) dR'= 2 \psi_{1}(R).
\ee
Similarly, for $\phi_{1}$ we arrive at the expression
\be
{d^2 \over dR^2}\phi_{1} + {1 \over R} \dR \phi_1 = 
{ \rho^2 \over R^2} \left (
{Q^2 \over r^4} \right ) [f_{0} - (\psi_{1} + \phi_{1} + \ga_{1})],
\ee
which implies that $ \phi_{1} = \tp \ln \sqrt{r(r - {Q^2 \over M})}
\sin \theta$, where $\tp$ is a constant value.
\par 
Taking into account the above corrections, one can consistently
transform the metric to the $(t, r, \theta, \ppp)$ coordinates in which
the asymptotic form of the metric is expressed as
\ben
ds^2 \rightarrow e^{\ep C} \left [ -
\left ( 1 - {2M \over r} \right ) dt^2 +
{d r^2 \over \left ( 1 - {2M \over r} \right ) } +
r \left ( r - {Q^2 \over M} \right ) d \theta^2 
\right ] + \\ \nonumber
+ r \left ( r - {Q^2 \over M} \right ) \left [ 1 - \ep A + {\ep B \over \sqrt {r ( r -
{Q^2
\over M})} \sin \theta }
\right ]^2 e^{- \ep C} \sin^2 \theta d\ppp^2,
\een
where $e^{\ep C} = e^{2 \ep \psi_{1}}$. \\
One should notice \cite{gg} that the $B$-term is outside the range of the
applicability of the considered approximation. After rescalling 
coordinates
$\hht = e^{ \ep C/2}t, \hhr = e^{\ep C/2} r$ and defining the quantities
$\hhM = e^{\ep C/2}M$ and $ \hhQ = e^{\ep C/2} Q$, 
one gets the metric in the {\it thin string} metric, i.e. $M \gg 1$,
\be 
ds^2 =  -
\left ( 1 - {2 \hhM \over \hhr} \right ) d \hht^2 + {d \hhr^2 \over 
\left ( 1 - {2 \hhM \over \hhr} \right ) } +
\hhr \left ( \hhr - {\hhQ^2 \over \hhM} \right ) d \theta^2 +
\hhr \left ( \hhr -{\hhQ^2 \over \hhM} \right ) (1- \ep A)^2 e^{-
2 \ep C} \sin^2 \theta
d\ppp^2 .
\ee
Now, we turn to a
deficit angle, which has the form
$\delta \theta = 2 \pi \ep (A + C) = 8 \pi G \mu$, where
$\mu $ is the total mass of string per unit length. On the
other hand, its ADM mass generalized to an
asymptotically flat space is written \cite{haw}
\be
M_{I} = \hhM (1 - \ep A) e^{- \ep C}.
\ee
The definition of the physical charge of the black hole,
respectively for magnetic or electric charge is given by
\be
Q_{ph} = {1 \over 4 \pi} \int_{S^2}
F_{\mu \nu} dS^{\mu \nu},
\qquad
Q_{ph} = {1 \over 4 \pi} \int_{S^2} \ast 
F_{\mu \nu} dS^{\mu \nu},
\ee
where $dS_{\mu \nu} = l_{[ \mu}n_{\nu ]} dA$ and 
$dA$ is the element of surface area. The null vector 
$n_{\mu}$ is
orthogonal to the 
two-sphere on the horizon, with the normalization condition
$l_{\alpha} n^{\alpha} = -1$. \\
Then, one can
write the first order corrected solutions in terms of $M_{I}$ and
$Q_{ph}$. The resultant metric has the form
\ben 
ds^2 = &-&
\left ( 1 - {2M_{I} e^{4 G \mu} \over \hhr} \right ) d \hht^2 +
{d \hhr^2 \over \left ( 1 - {2M_{I} e^{4 G \mu} \over \hhr} \right ) } +
\hhr \left ( \hhr - {Q_{ph}^2 e^{4 G \mu - 2 \ep \tD}
\over M_{I}} \right ) d \theta^2 \\ \nonumber
&+&
\hhr \left (
\hhr -{Q_{ph}^2 e^{4 G \mu - 2 \ep \tD}
\over M_{I}} \right ) e^{- 4 G \mu} \sin^2 \theta
d\ppp^2, 
\een
where $\tD = 2 \phi_{1} + f_{0}$. The corrected inner $\hhr_{-}$
and outer $\hhr_{+}$ horizons are situated at
\be
\hhr_{-} = {Q_{ph}^2 e^{4 G \mu - 2 \ep \tD} \over M_{I}}, \qquad
\hhr_{+} = 2 M_{I} e^{4 G \mu}.
\ee
On the other hand, the corrected condition for the extremal black hole
yields 
\be
2 M_{I}^2 = Q_{ph}^2 e^{- 2 \ep \tD}.
\ee
Finally, using the formula for the entropy \cite{haw}, 
we conclude that it has form as follows:
\be
S = 2 \pi M_{I} e^{6 G \mu} \left (
2 M_{I} - {Q_{ph} e^{-2 \ep \tD} \over M_{I}} 
\right ).
\ee

\vspace{0.3cm}
Finally, we will discuss
another interesting problem that can be explored, i.e.,
the interaction of
a cosmic string with an extreme dilatonic black hole. 
We shall
consider the
vortex which is in a perfect aligment with an extremal black hole axis.
This assumptions enables one to get rid of great complications when the
black hole and the vortex are displaced to each other.
We also assume that, we do not take into account
details of the core structure
of a cosmic string. 
It is turned out that the extreme dilatonic black hole metric can be
written as \cite{sir1}
\ben
ds^2 = - {1 \over V}dt^2 + V(dx^2 + dy^2 + dz^2) \\ \nonumber
A_{\mu} = {1 \over \sqrt{2}} \left ( 1 - {1 \over V} \right ) \p_{\mu}t,
\een
where the function 
$V$ satisfies ${\na^2}_{(x, y, z)} V = 0$. In the cylindrical
coordinates $(\rho, z, \varphi)$ centered on the string, with a conical
deficit $0 \leq \varphi \leq {2 \pi\over p}$, 
$\enskip p \approx 1 + 4 \mu G$,
the function $V$ takes the
form (see, e.g., Ref.\cite{ss}) 
\be
V(z, \rho, \varphi; \rho_{0}) =
1 + {2 M \over \pi \sqrt{2 \rho \rho_{0}}}
\int_{u_{0}}^{\infty} {du \over \sqrt{\cosh u - \cosh u_{0}}}
{p \sinh pu \over \cosh pu - \cos p \varphi},
\ee
where we set the black hole at $\rho = \rho_{0}, \varphi = 0$, and $z =
0$. While $ u_{0}$ is defined by the relation $\cosh u_{0} =
{\rho^2 + z^2 + \rho_{0}^2 \over 2 \rho \rho_{0}}$. The obtained
function is nonsingular away from the conical line and the singularity of
the black hole. Then, there are no forces between the extremal black
hole and the cosmic string. Analogous results have been obtained in
the case of a cosmic string and an extremal 
Reissner-Nordstr\"om black hole
\cite{bon}.
Of course, we should be aware of neglecting the effect of the dilaton
extremal black hole on the string core. Nevertheless,
using our assumption of a
complete separation of degrees of freedom of each of the objects
one concludes
that an extremal dilaton black hole and a straight cosmic string
will hardly feel their presence.

\section{Conclusions}
In our work we ask the question of whether or not an Abelian-Higgs
vortex is expelled from the extremal dilaton black hole. We gave
analytical arguments that no matter how thick was the vortex it 
was always expelled from the considered black hole.
In order to confirm our analytic results we performed numerical
calculations in which boundary conditions on the horizon 
of the extreme dilatonic black hole were guessed at the beginning of the
process and updated according to the adequate equations. 
We also paid attention to the vortices ending on the extremal
dilaton black hole.
\par
Finally, we studied the backreaction effect of the vortex on the
geometry and the other fields in the theory under consideration.
In {\it the thin string} limit we get the {\it conical} dilaton black
hole metric. Concluding, we mentioned the problem of an interaction
between a straight cosmic string and an extremal dilaton black hole
which was situated in the perfect aligment with a black hole axis.
According to the assumptions of the clear separation of
the degrees of freedom of these objects, one can conclude
that they hardly feel their
presence.

\vspace{3cm}
\noindent
{\bf Acknowledgements:}\\
We would like to thank Ruth Gregory and Roberto Emparan 
for helpful remarks and discussions on various
occasions. R.M. acknowledges support from the NASA Long Term
Astrophysics grant NASA-NAG-6337 and NSF grant No. AST-9529175.


\pagebreak

\begin{figure}[p]
\begin{center}
\leavevmode
\epsfxsize=440pt
\epsfysize=440pt
\epsfbox{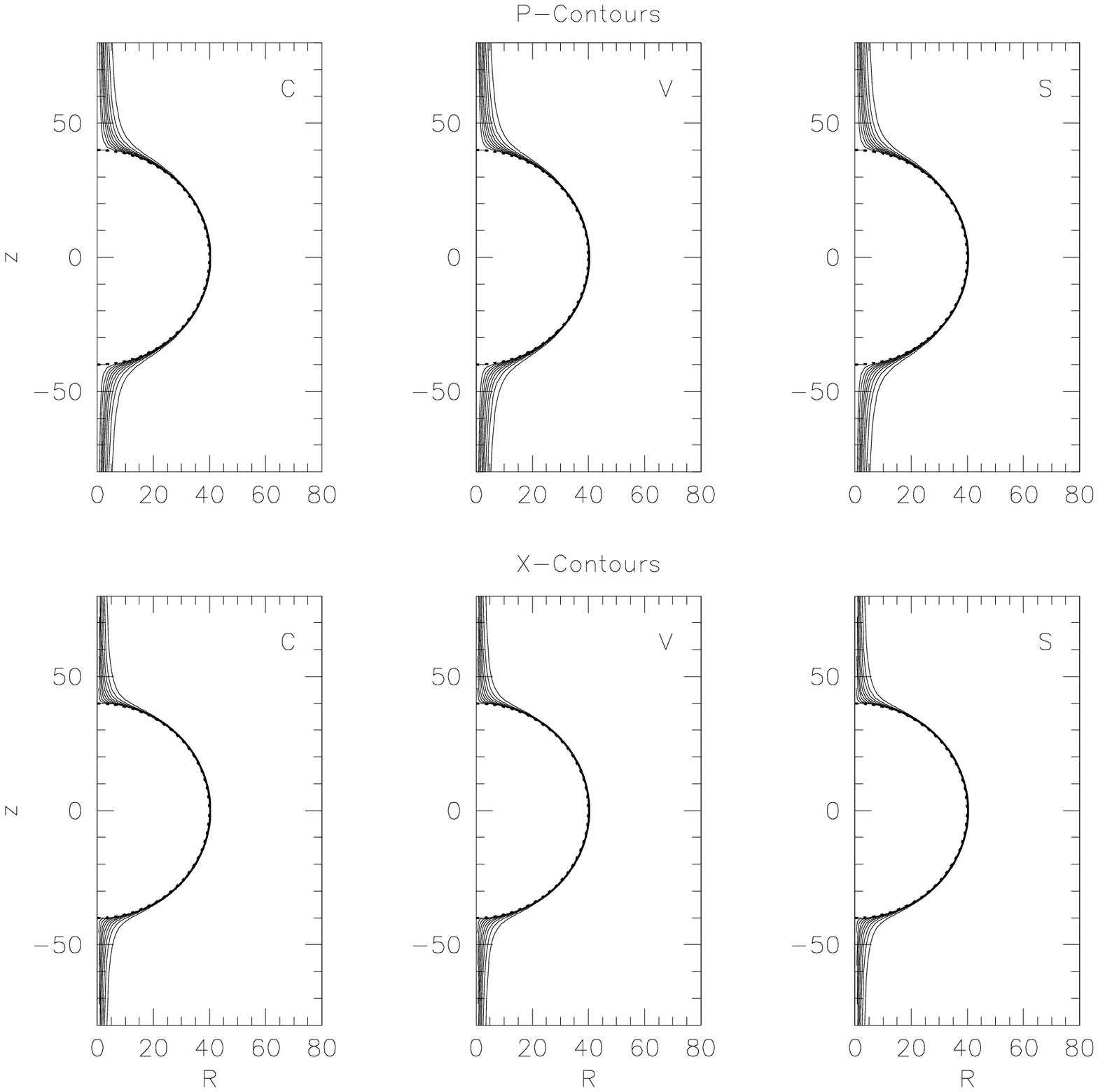}
\end{center}
\caption{\bf Contours of $X$ and $P$ for a core (C), vacuum (V), 
and sinusoidal (S)
initial guess. The parameters $M = 1, Q = 1.41 M, \beta = N = 1$.}
\label{fig1}
\end{figure}


\begin{figure}[p]
\begin{center}
\leavevmode
\epsfxsize=440pt
\epsfysize=440pt
\epsfbox{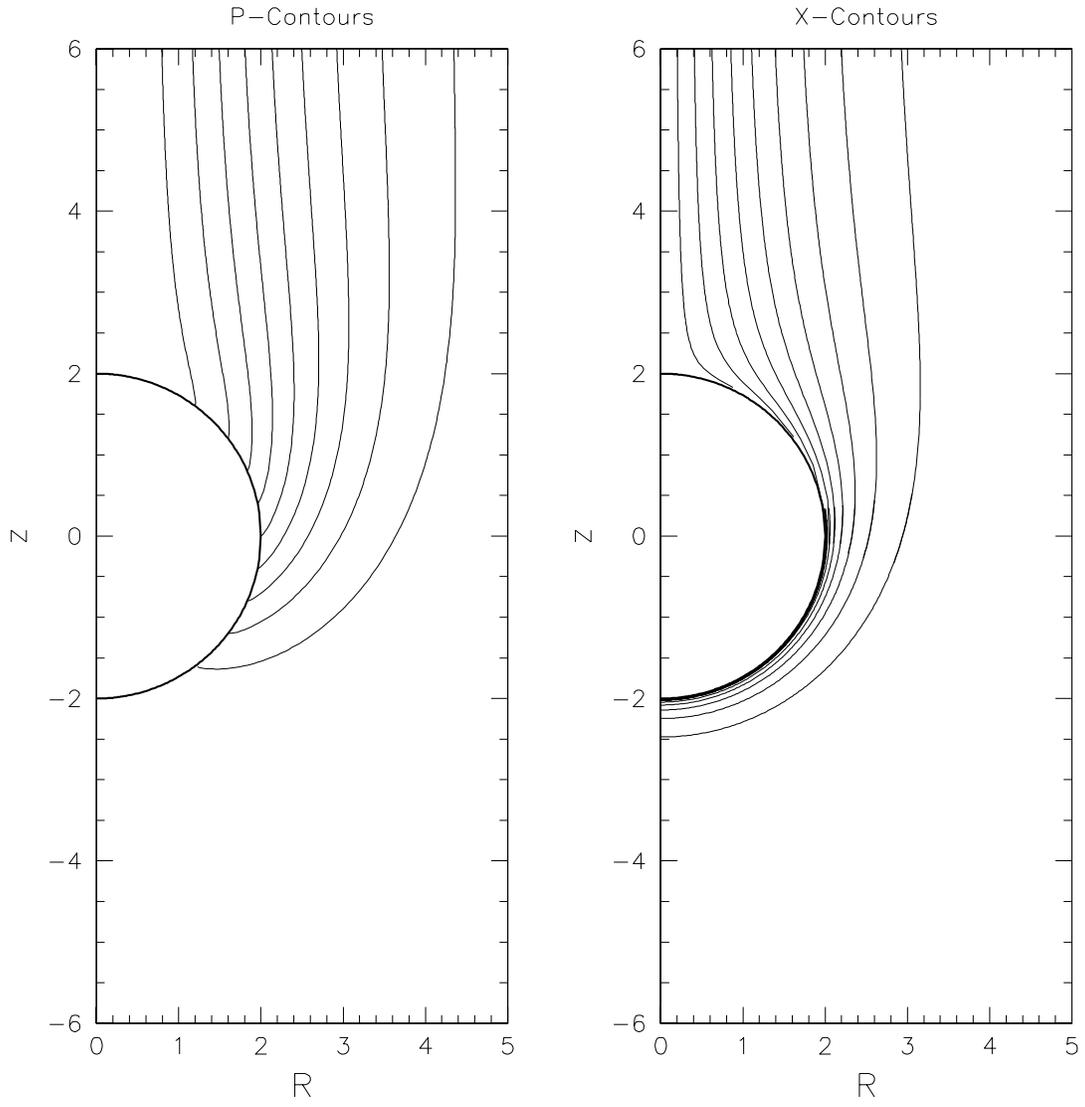}
\end{center}

\caption{\bf Contours of $X$ and $P$ for a 
single string ending on the extremal dilaton black hole.
The parameters $M = 1, Q = 1.41 M, \beta = N = 1$.}
\label{fig2}
\end{figure}
\end{document}